\documentclass[aps,superscriptaddress,floatfix,reprint]{revtex4-1}
\usepackage{amsmath,amssymb}
\usepackage{color}
\usepackage[colorlinks=true,urlcolor=blue]{hyperref}
\usepackage{bm}
\usepackage{graphicx}
\usepackage[english]{babel}
\usepackage{siunitx}

\begin{document}

\title{Effective self-similar expansion of a Bose-Einstein condensate:
\\
Free space versus confined geometries}

\author{David Viedma}
\affiliation{Department of Theoretical Physics and History of Science, University of the Basque Country UPV/EHU, 48080 Bilbao, Spain}
\author{Michele Modugno}
\affiliation{Department of Theoretical Physics and History of Science, University of the Basque Country UPV/EHU, 48080 Bilbao, Spain}
\affiliation{IKERBASQUE, Basque Foundation for Science, 48013 Bilbao, Spain}

\date{\today}

\begin{abstract}
We compare the exact evolution of an expanding three-dimensional Bose-Einstein condensate with that obtained from the effective scaling approach introduced in D. Gu\'ery-Odelin [Phys. Rev. A \textbf{66}, 033613 (2002)]. 
This approach, which consists in looking for self-similar solutions to be satisfied on average, is tested here in different geometries and configurations.
We find that, in case of almost isotropic traps, the effective scaling reproduces with high accuracy the exact evolution dictated by the Gross-Pitaevskii equation for arbitrary values of the interactions, in agreement with the proof-of-concept of M. Modugno, G. Pagnini, and M. A. Valle-Basagoiti [Phys. Rev. A \textbf{97}, 043604 (2018)]. Conversely, it is shown that the hypothesis of universal self-similarity breaks down in case of strong anisotropies and trapped geometries. 
\end{abstract}
\maketitle

\section{Introduction}

Self-similarity is a remarkable property that plays a key role for describing the dynamics of several ultracold atomic systems. It characterizes the free expansion and the collective excitations of Bose-Einstein condensates both in the noninteracting and the hydrodynamic regimes \cite{castin1996,kagan1996,dalfovo1999,timmermans2000,brazhnyi2003,kamchatnov2004,egusquiza2011,quach2014,li2019}, the expansion of a one-dimensional Bose gas in the mean-field Thomas-Fermi regime and in the Tonks-Girardeau regime \cite{pedri2003}, of a superfluid Fermi gas \cite{giorgini2008,schafer2010,egusquiza2011}, and of a thermal cloud \cite{bruun2000}. 
Recently, motivated by the experiment reported in Ref. \cite{chang2016}, scaling transformations have been employed for understanding how the momentum distribution is affected by the expansion in an interacting quantum system \cite{qu2016}, and the conditions for the breaking of scale invariance have also been investigated \cite{gharashi2016}.

Beside its conceptual appeal -- the fact that a system evolves maintaining the same exact shape throughout the whole dynamics --, self-similarity also implies a strong simplification in the numerical treatment, allowing for a dramatic reduction of the mathematical complexity of the equation governing the system: Instead of having to deal with partial differential equations, one can obtain the evolution of the system by solving a set of ordinary differential equations for the time-evolution of the scaling parameters that characterize the rescaling of the coordinates. 

In view of this numerical simplification, approximate self-similar behaviors have been conjectured by several authors for dealing with problems that otherwise would be impossible to tackle. Effective scaling approaches have been employed as approximate solutions for describing the collective excitations of a trapped Bose gas \cite{guery-odelin2002}, the expansion of an interacting Fermi gas \cite{menotti2002}, and of quantum degenerate Bose-Fermi mixtures \cite{hu2003}. This approach, originally proposed in Ref.~\cite{guery-odelin2002}, consists in using a self-similar ansatz for the evolution of the system in the hydrodynamic regime, then requiring the hydrodynamic equations to be satisfied \textit{on average}, after integration over the spatial coordinates.

Recently, this effective approach was tested for the case of a freely expanding quasi-one-dimensional Bose-Einstein condensate (BEC), by comparing the approximate solution with the exact evolution of the system as obtained from the solution of the Gross-Pitaevskii (GP) equation. Remarkably, in this case it was found that the effective scaling approach is indeed very accurate in reproducing the exact evolution, for arbitrary values of the interactions \cite{modugno2018}. 

In the present paper we extend this analysis to the case of a three-dimensional (3D) BEC \footnote{We recall that for a harmonically trapped 2D condensate self-similar solutions are universal \cite{kagan1996}.}, for which we consider the expansion in free space and in trapped geometries (namely, in cylindrical and planar waveguides). 
The aim of this work to understand in a quantitative framework how nonlinearity affects self-similar dynamics, 
in the context of the GP equation, for which scaling approaches have been widely employed \cite{castin1996,kagan1996,dalfovo1999,timmermans2000,brazhnyi2003,kamchatnov2004,egusquiza2011,modugno2018,li2019}.
This is particularly relevant in 3D -- where the interplay between spatial anisotropies and interactions may not be trivial -- and where there are not actual proofs that justify the usage of an \textit{effective} scaling approach.
Moreover, since time-of-flight expansion is still a very relevant part of current experiments with BECs, as e.g. those with binary mixtures \cite{lee2018,burchianti2018}, a quantitative assessment on the validity of this approach may be very useful for extending such a method to more complex configurations.

In case of the expansion in free space, we find that the effective scaling approach for the expansion of a spherically symmetric condensate is rather accurate even for intermediate values of the interaction -- in between the noninteracting and hydrodynamic limits where the scaling is exact --, similarly to what found for the quasi-1D case \cite{modugno2018}. Deviations from this behavior are observed instead for prolate and especially oblate condensates, signalling that the scaling approach becomes less accurate when the expansion along certain directions is faster than along the others, causing local variations of the density that do not conform to the hypothesis of self-similarity.
Likewise, self-similarity may also break down in the presence of a residual trapping. In fact, the density profile along the trapped directions is determined by the interplay of the confining potential and the nonlinear term, and it therefore gets modified as the system expands.
In this scenario, the scaling approach can be safely employed only in the noninteracting limit (for obvious reasons) and in the hydrodynamic limit. In the latter case, a necessary condition is also that the evolution time should be sufficiently short such that the initial Thomas-Fermi (TF) profile along the trapped directions is preserved. This is conceptually different from the case of the evolution in free space, where e.g. once a TF profile is fixed by the initial conditions it is then maintained during the whole evolution, regardless of the variation in the local value of the nonlinear term. 

The paper is organized as follows. In Sec.~\ref{sec:scaling} we review the general procedure for constructing the effective scaling approach within the hydrodynamic formulation of the GP equation, as well as the form of the GP equation in the rescaled coordinate system. Then, in Sec.~\ref{sec:free} we apply this approach to the case of a freely-expanding condensate in 3D, considering first the case of a spherically symmetric BEC, and then the cases of oblate and prolate geometries. In Sec.~\ref{sec:cylindrical}, as an example of confined geometries, we examine the expansion in a waveguide and also comment on the case of a planar one. 
Final considerations are drawn in the conclusions.

\section{Scaling approach}
\label{sec:scaling}

Let us consider a Bose-Einstein condensate described by a wave function $\psi(\bm{x},t)$ that evolves according to the GP equation \cite{dalfovo1999} 
\begin{equation}
i\hbar \frac{\partial}{\partial t} \psi = \left[-\frac{\hbar^2}{2m} \nabla^2 + V(\bm{x},t) + g|\psi|^2\right]\psi,
\label{eq:GPE}
\end{equation}
with $m$ being the particle mass, $g=4\pi\hbar^2a/m$ the interaction strength (with $a$ the $s$-wave scattering length), and $V(\bm{x},t)$ a generic trapping potential, that may depend on time. In particular, we shall consider the case of a time-dependent harmonic potential of the form
\begin{equation} 
V(\bm{x},t)=\frac12 m\sum_i\omega_i^2(t)x_i^2,
\end{equation}
and we shall focus on the expansion dynamics of the condensate following the release of the confinement along some direction.
Equation (\ref{eq:GPE}) can be transformed into a system of two coupled hydrodynamic equations by means of a Madelung transformation, $\psi(\bm{x},t)\equiv\sqrt{n(\bm{x},t)}\exp{\{iS(\bm{x},t)\}}$ \cite{dalfovo1999},
\begin{gather}
\partial_t n + \bm{\nabla}\cdot(n\bm{v})=0,
\label{eq:2-hydro1}\\
m\partial_t \bm{v} + \bm{\nabla}\left(P+\frac{1}{2}mv^2 + V + gn\right)=0,
\label{eq:2-hydro2}
\end{gather}
where $\bm{v}\equiv(\hbar/m)\bm{\nabla} S$, and
\begin{equation}
P(\bm{x},t) = -\frac{\hbar^2}{2m}\frac{\nabla^2\sqrt{n}}{\sqrt{n}}
\label{eq:quantump}
\end{equation}
represents the so-called \textit{quantum pressure} term.

\textit{Exact Scaling.}
The scaling approach consists in looking for solutions characterized by a density profile that evolves \textit{self-similarly} as 
\begin{equation}
n(\bm{x},t) = \frac{1}{\prod_{j}{\lambda_j(t)}}n_{0}\left(\frac{x_{i}}{\lambda_i(t)}\right),
\label{eq:scaling-n}
\end{equation}
with $n_{0}(\bm{x})$ being fixed by the initial conditions, and with all the time dependence 
being contained within the scaling parameters $\lambda_i(t)$ (here the term $\prod_j{\lambda_j^{-1}}$ is just a volume normalization factor, with $i,j=1,2,3$ labeling the spatial directions). 
In addition, the continuity equation (\ref{eq:2-hydro1}) yields the following scaling of the velocity field: $v_i(\bm{x},t) = x_i{\dot{\lambda}_i(t)}/{\lambda_i(t)}$.

By introducing the ansatz (\ref{eq:scaling-n}) into the hydrodynamic equation (\ref{eq:2-hydro2}), one gets
\begin{equation}
m\frac{\ddot{\lambda}_i}{\lambda_i}x_i + \nabla_i\left[P + V + \frac{gn_{0}({x_{i}}/{\lambda_i(t)})}{\prod_j{\lambda_j}}\right] = 0.
\label{eq:scaling-v}
\end{equation}
In the noninteracting regime ($g=0$) and in the so-called hydrodynamic regime (negligible $P$) the above expression can be factorized in two terms, one depending on the spatial coordinates and the other on the time coordinate alone, so that the scaling ansatz (\ref{eq:scaling-n}) represents an \textit{exact} solution.

\textit{Effective Scaling.}
 Contrarily, when such a factorization is not possible, one may follow a different approach by requiring the self-similarity to be satisfied on average, as discussed in Refs. \cite{guery-odelin2002,menotti2002,hu2003,modugno2018}.
First, it is convenient to transform the expression in Eq.~(\ref{eq:scaling-v}) by integrating over the $i$-th coordinate \footnote{This is an \textit{indefinite integration}, not to be confused with the averaging procedure that will be used afterwards.},
\begin{equation}
\frac{m}{2}\frac{\ddot{\lambda}_i}{\lambda_i}x_i^2 + P + V + \frac{gn_{0}({x_{i}}/{\lambda_i(t)})}{\prod_j{\lambda_j}} - q(\bm{x},t)|_{x_i=0} = 0,
\label{eq:3-sc1}
\end{equation}
where $q(\bm{x},t) \equiv P(\bm{x},t) + V(\bm{x},t) + {gn_{0}({x_{i}}/{\lambda_i(t)})}/{\prod_j\lambda_j}$.
Then, we multiply Eq.~(\ref{eq:3-sc1}) by $n_{0}({x_{i}}/{\lambda_i(t)})$ and we get rid of the coordinate dependence by integrating over the volume.
This sort of averaging procedure -- that constitutes the essence of the \textit{effective scaling} approach \cite{guery-odelin2002,modugno2018} -- yields an equation for the scaling parameters depending only on the time variable. 
Specific expressions for each case considered in this paper will be discussed in the following sections.

\textit{Scaled Gross-Pitaevskii equation.}
The scope of this work is to test the accuracy of the effective scaling approach in reproducing the exact evolution by quantifying the deviation of the scaling dynamics from the exact solution of the GP equation in (\ref{eq:GPE}).
To this end, it is convenient to rewrite the GP equation in terms of the rescaled wave function, defined as \cite{castin1996}
\begin{equation} 
\label{eq:scaling-psi}
\psi(\bm{x},t)=\frac{1}{\sqrt{\prod_j\lambda_j(t)}}\phi\left(\frac{x_{i}}{\lambda_{i}(t)},t\right)e^{\displaystyle i\frac{m}{2\hbar}\sum_{j}\frac{\dot\lambda_{j}(t)}{\lambda_{j}(t)}x_{j}^{2}},
\end{equation}
with $\phi(x,0)=\sqrt{n_{0}(x)}$. Then, plugging the above expression for $\psi(x,t)$ in
the Gross-Pitaevskii equation (\ref{eq:GPE}) yields 
\begin{align}
\label{eq:scaling-gp}
&i\hbar\partial_{t}\phi=
 -\frac{\hbar^2}{2m}\sum_{j}\frac{1}{\lambda_{j}^{2}}\widetilde{\nabla}_{j}^{2}\phi 
 \\
 &\qquad+\frac{1}{2}m\sum_{j}
 \left(\omega_j^2(t)+\frac{\ddot\lambda_{j}}{\lambda_{j}}\right)x_{j}^{2}\phi+\frac{g}{\prod\lambda}|\phi|^{2}\phi,
 \nonumber
\end{align}
where we have defined $\widetilde{\nabla}_{j}=\partial/\partial(x_{j}/\lambda_{j})$. Notice that the partial derivative $\partial_t$ only operates on the explicit dependence on time in $\phi(\cdot,t)$,
which accounts only for the deviation from the self-similar solution.
The latter can be measured by defining the following \textit{fidelity} \cite{modugno2018}
\begin{equation}
 F(t)\equiv|\langle\phi(0)|\phi(t)\rangle|,
\label{eq:fidelity}
\end{equation}
that can take values in the range $F\in[0,1]$, the upper bound ($F=1$) corresponding to exact self-similar solutions.
This quantity provides a quantitative estimate of the accuracy of the effective approach.

In the following, it is also useful to introduce dimensionless units, which can be conveniently defined in terms of natural oscillator units of a suitable frequency $\omega_{0}$, namely $\tau\equiv\omega_{0}t$, 
with the spatial coordinates being expressed in units of
$a_{ho}\equiv\sqrt{\hbar/(m\omega_{0})}$, and so on. As for the interaction strength, we define $\tilde{g}\equiv 4\pi N a/a_{ho}$, by including also the number of atoms $N$ in the definition. From now on, all the quantities will be considered as dimensionless, unless otherwise stated.

\section{Expansion in free space} 
\label{sec:free}

Here we consider a condensate that expands in free space, after being released from a three-dimensional harmonic trap. For the sake of simplicity, we restrict the analysis to axially symmetric potentials, namely
\begin{equation}
 V(r_{\perp},z) = \frac12\left[\omega_{\perp}^2(\tau)r_{\perp}^2 + \omega_z^2(\tau)z^2\right],
 \label{eq:cylpot}
\end{equation}
 where $\omega_{\perp,z}(\tau)\equiv0$ for $\tau>0$. 
As unit frequency we choose the geometric average of the trapping frequencies in the three spatial directions $\omega_{0}=\sqrt[3]{\omega_{\perp}^2(0)\omega_{z}(0)}$,
and we define the parameter $\alpha=\omega_{\perp}(0)/\omega_z(0)$ characterizing the trap anisotropy.
In this case, the self-similar ansatz for the density takes the following form
\begin{equation}
n(r_{\perp},z,\tau) = \frac{1}{\lambda_{\perp}^2(\tau)\lambda_z(\tau)}
n_{0}\left(\frac{r_{\perp}}{\lambda_{\perp}(\tau)},\frac{z}{\lambda_z(\tau)}\right).
\label{eq:scaling-cyl}
\end{equation}
The corresponding scaling equations are (see Appendix for the derivation and the definition of the various constants)
\begin{align}
\ddot{\lambda}_{\perp} &= \frac{A_{1\perp}(\tilde{g})}{\lambda_{\perp}^3} + \frac{A_{2\perp}(\tilde{g})}{\lambda_{\perp}\lambda_z^2} + \frac{B_{\perp}(\tilde{g})}{\lambda_{\perp}^3\lambda_z}, 
\label{eq:lrf} 
\\
\ddot{\lambda}_z &= \frac{A_{1z}(\tilde{g})}{\lambda_{\perp}^2\lambda_z} + \frac{A_{2z}(\tilde{g})}{\lambda_z^3} + \frac{B_{z}(\tilde{g})}{\lambda_{\perp}^2\lambda_z^2},
\label{eq:lzf}
\end{align}
where the constants $A_{i\nu}$ and $B_{\nu}$ ($i=1,2$, $\nu=\perp,z$) -- which are fixed by the initial conditions -- fulfill the following sum rules
\begin{equation}
\sum_{i=1,2}A_{i\perp} + B_{\perp} = \alpha^{2/3};\;
\sum_{i=1,2}A_{iz} + B_{z} = \alpha^{-4/3}.
\label{eq:sumrules}
\end{equation}

If the initial condensate is spherically symmetric, $\omega_{\perp}=\omega_{z}\equiv \omega$ ($\alpha=1$), then the above Eqs.~(\ref{eq:lrf}) and (\ref{eq:lzf}) become degenerate, and obviously they greatly simplify. In fact, the system can be described in terms of a radial coordinate alone, $r\equiv\sqrt{r_{\perp}^{2}+z^{2}}$, and it is characterized by a single scaling parameter $\lambda(\tau)$.
The above equations simplify to
\begin{equation}
\ddot{\lambda}(\tau) = \frac{A(\tilde{g})}{\lambda^3(\tau)} + \frac{B(\tilde{g})}{\lambda^4(\tau)},
\label{eq:scalingsph}
\end{equation}
where $A=\sum_{i=1,2}A_{i\nu}$ and $B=B_{\nu}$ (any $\nu$), see the Appendix. 

\begin{figure}
\centering
 \includegraphics[width=0.9\columnwidth]{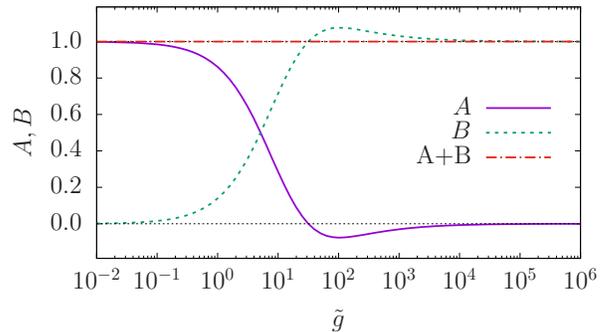}
\caption{Behavior of the constants of the scaling equation (\ref{eq:scalingsph}) as a function of the interaction strength $\tilde{g}$.}
\label{fig:AB}
\end{figure}

Manifestly, the above formulas in Eqs.~(\ref{eq:lrf})-(\ref{eq:lzf}) and (\ref{eq:scalingsph}) interpolate between the noninteracting ($\lambda^{-3}$ terms) and TF ($\lambda^{-4}$ term) regimes. 
The behavior of the coefficients $A$ and $B$ as a function of the interaction strength $\tilde{g}$, see Fig.~\ref{fig:AB}, indeed guarantees that the known results \cite{dalfovo1999} corresponding to the exact solutions in the noninteracting ($\tilde{g}\ll1$) and Thomas Fermi limits ($\tilde{g}\gg1$) are recovered. A similar behavior is displayed by the parameters $A_{i\nu}$ and $B_{\nu}$ of the axially-symmetric case, see the Appendix.

In the following, we shall first analyze to which extent the hypothesis of effective self-similarity is able to capture the actual GP dynamics for a spherically symmetric condensate, as a function of the interaction strength $\tilde{g}$, that is the only parameter the system depends on. Then, we discuss the effect of the trap anisotropy $\alpha$, by extending the analysis to a generic axially symmetric condensate.

\begin{figure}
\centering
 \includegraphics[width=0.9\columnwidth]{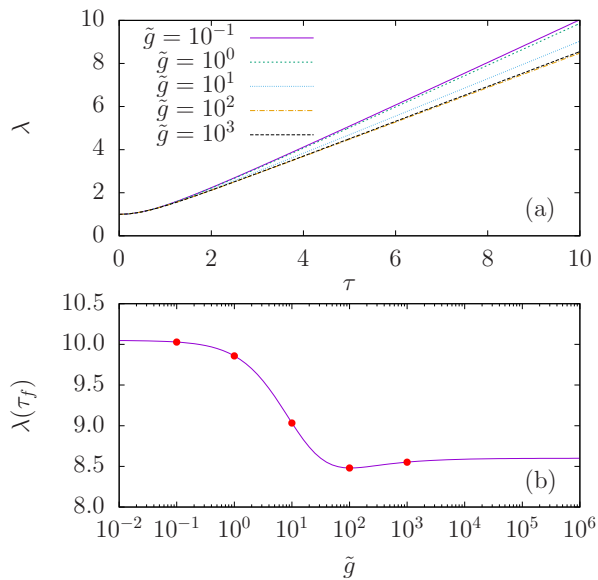}
\caption{Behavior of the scaling parameter $\lambda(\tau)$ (a) as a function of time for different interaction strengths and (b) at $\tau_{f}=10$ as a function of the interaction strength. The red dots in (b) correspond to the values of $\tilde{g}$ considered in (a).}
\label{fig:lambda-sph}
\end{figure}

\begin{figure}
\centerline{
 \includegraphics[width=0.9\columnwidth]{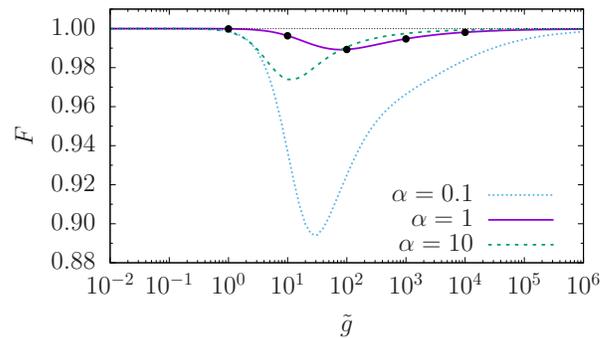}}
\caption{Behavior of the fidelity as a function of the interaction strength $\tilde{g}$, at $\tau_{f}=10$, for different values of the trap anisotropy $\alpha$. The black circles correspond to the density profiles shown in Fig.~\ref{fig:psir-sph}.}
\label{fig:fidelity}
\end{figure}

\textit{Spherical condensate.}
The behavior of the scaling parameter $\lambda(\tau)$ as a function of time (for different values of $\tilde{g}$), and its final value $\lambda_{f}\equiv\lambda(\tau_{f})$ ($\tau_{f}=10$) as a function of $\tilde{g}$, are shown in Figs.~\ref{fig:lambda-sph}(a) and (b), respectively. 
Remarkably, the final value $\lambda_{f}(\tilde{g})$ decreases by increasing the nonlinear coupling, and it has a slight minimum at the crossover region between the noninteracting and hydrodynamic regimes.
The first behavior is directly connected to the power law in the scaling equation (\ref{eq:scalingsph}): In the TF limit the right-hand side behaves like $\propto\lambda^{-4}$ \cite{li2019}, and this implies a slower growth of $\lambda$ ($>1$) as compared to the $\propto\lambda^{-3}$ behavior of the noninteracting limit. This reflects the fact that as $\tilde{g}$ is increased, the density distribution gets wider and this narrows the momentum distribution, thus reducing the contribution of kinetic energy to the condensate expansion. As for the nonmonotonic behavior at intermediate values of $\tilde{g}$, we notice that the minimum of $\lambda_{f}$ corresponds to the minimum of the fidelity, shown in Fig.~\ref{fig:fidelity}. 
This figure shows that the accuracy of the effective scaling can be very high, with a fidelity always above $98\%$, for any value of the interaction strength ($\tau \leq {10}$). This approach seems therefore quite robust, similarly to the 1D case discussed in Ref.~\cite{modugno2018}.
\begin{figure}
\centering
 \includegraphics[width=0.9\columnwidth]{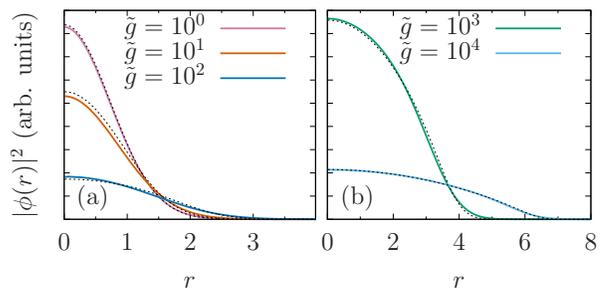}
\caption{Comparison of the radial density profiles of a spherical condensate obtained from the numerical solution of the full GP equation (solid) and from the scaling argument (dashed) for different interaction strengths, at the final evolution time $\tau_{f}=10$ (the corresponding fidelity is marked by the black circles in Fig.~\ref{fig:fidelity}). 
The vertical axes on the two panels are not on scale.}
\label{fig:psir-sph}
\end{figure}

Its reliability can also be appreciated from the behavior of the density profiles, which are shown in Fig.~\ref{fig:psir-sph} for different values of $\tilde{g}$. Only small deviations can be seen, especially around the minimum of fidelity (at $\tilde{g}\equiv \tilde{g}^{*}$). It is also interesting to notice that for $\tilde{g}<\tilde{g}^{*}$ the scaling ansatz produces a slightly larger central value for the density, whereas the opposite happens for $\tilde{g}>\tilde{g}^{*}$.

\begin{figure}
\centerline{\includegraphics[width=0.98\columnwidth]{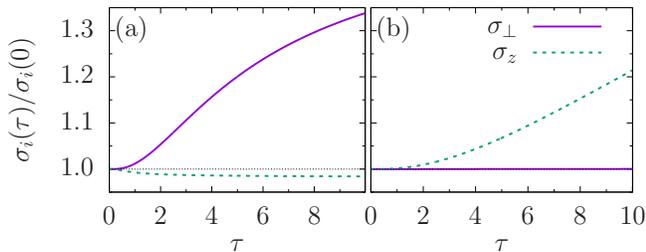}}
\caption{Evolution of the axial and radial widths, normalized to the corresponding initial values, $\sigma_{z}(\tau)/\sigma_{z}(0)$ and $\sigma_{\perp}(\tau)/\sigma_{\perp}(0)$, respectively. (a) Oblate geometry, $\alpha=0.1$ (pancake); (b) prolate geometry, $\alpha=10$ (cigar). Here $\tilde{g}=10$.}
\label{fig:sigmas}
\end{figure}

\textit{Pancake- and cigar-shaped condensates.} In the following, we shall compare the case of oblate and prolate geometries, with $\alpha=10$ and $\alpha=0.1$ respectively, with the previously considered spherical case ($\alpha=1$). The corresponding values of the fidelity after an expansion time $\tau_{f}=10$ are shown in Fig.~\ref{fig:fidelity}. This figure shows that, in the presence of a trap anisotropy, the assumption of a self-similar evolution is less effective for intermediate values of the interactions, especially for the case of oblate geometries (the case with $\alpha=0.1$ in the figure). A qualitative argument for explaining this behavior is the following. 
When a pancake-shaped condensate is released from the trap, the initial expansion is characterized by a fast dilatation along the axial direction, whereas the radial dynamics is much slower. The former causes a fast local variation of the density, which drives the radial expansion out of the self-similar regime. This happens due to the fact that the radial dynamics soon decouples from that of the mean-field term, contrarily to what occurs e.g. in the expansion of a spherical condensate. This is confirmed by Fig.~\ref{fig:sigmas}(a), where we plot the behavior as a function of time of the condensate radial and axial widths, $\sigma_{\perp}(\tau)\equiv\sqrt{\langle r_{\perp}^{2}\rangle_{\tau}}$
and $\sigma_{z}(\tau)\equiv\sqrt{\langle z^{2}\rangle_{\tau}}$, respectively. 
This figure shows that the value of $\sigma_{z}(\tau)/\sigma_{z}(0)$ remains very close to $1$ in the rescaled coordinate system we are using, indicating that the evolution of the condensate along the axial direction is well reproduced by the hypothesis of self-similarity. However, the expansion in the transverse plane is faster than what is dictated by the self-similar approach, signaling that the latter substantially underestimates the actual value of $\sigma_{\perp}(\tau)/\sigma_{\perp}(0)$.
\begin{figure}
\centerline{
 \includegraphics[width=0.9\columnwidth]{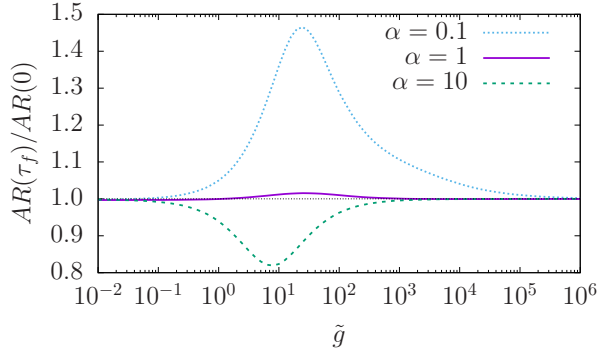}}
\caption{Rescaled aspect ratio as a function of the interaction strength $\tilde{g}$, at $\tau_{f}=10$, for different values of the trap anisotropy $\alpha$. The lines represent the exact values obtained from the solution of the rescaled GP equation (\ref{eq:scaling-gp}), whereas the dotted horizontal line correspond to the ideal value $AR(\tau_{f})/AR(0)\equiv 1$ of an exact self-similar expansion.}
\label{fig:ar}
\end{figure}

In the opposite case of a cigar-shaped condensate, the fidelity substantially improves (though not at the level of the spherically symmetric case), see the curve for $\alpha=10$ in Fig.~\ref{fig:fidelity}. The reason for this is that here the dynamics is driven by two of the three spatial directions, so that the breaking of the self-similarity mostly affects one direction only, namely the axial one. Indeed, Fig.~\ref{fig:sigmas}(b) shows that the scaling for the transverse width is almost exact, whereas the axial width increases faster that what is predicted by the self-similar expansion. This confirms that the self-similar approach underestimates the expansion along the ``slow'' direction.

The above behavior has a direct consequence on the value of the aspect ratio $AR(\tau)\equiv\sigma_{\perp}(\tau)/\sigma_{z}(\tau)$, a quantity that is typically used to characterize the condensate expansion \cite{dalfovo1999,menotti2002,hu2003}. Consistently with the scaling approach, here it is convenient to consider the ratio $AR(\tau)/AR(0)$. 
In case of an exact self-similar expansion, the former quantity is expected to be constant and equal to one in the rescaled coordinate system we are using, $AR(\tau)/AR(0)\equiv 1$ for any $\tau$. The deviations from the ideal behavior are shown in Fig.~\ref{fig:ar}. We see that in the spherical case the self-similar approach provides very accurate results, whereas in case of oblate and prolate condensates significant deviations are observed for intermediate values of the interaction (up to $50\%$), as one may expect from what was discussed previously.

To conclude this section, we remark that the scaling approach can be still very useful from the numerical point of view, even when it is not able to reproduce the exact evolution. Indeed, it allows us to replace the usual GP equation with the one in Eq.~(\ref{eq:scaling-gp}), which evolves the system in a rescaled coordinate system, thus greatly reducing
the size of the computational grid in case of a free expansion.

\section{Confined geometry: Expansion in a waveguide} 
\label{sec:cylindrical}

\begin{figure*}
\centering
 \includegraphics[width=0.9\columnwidth]{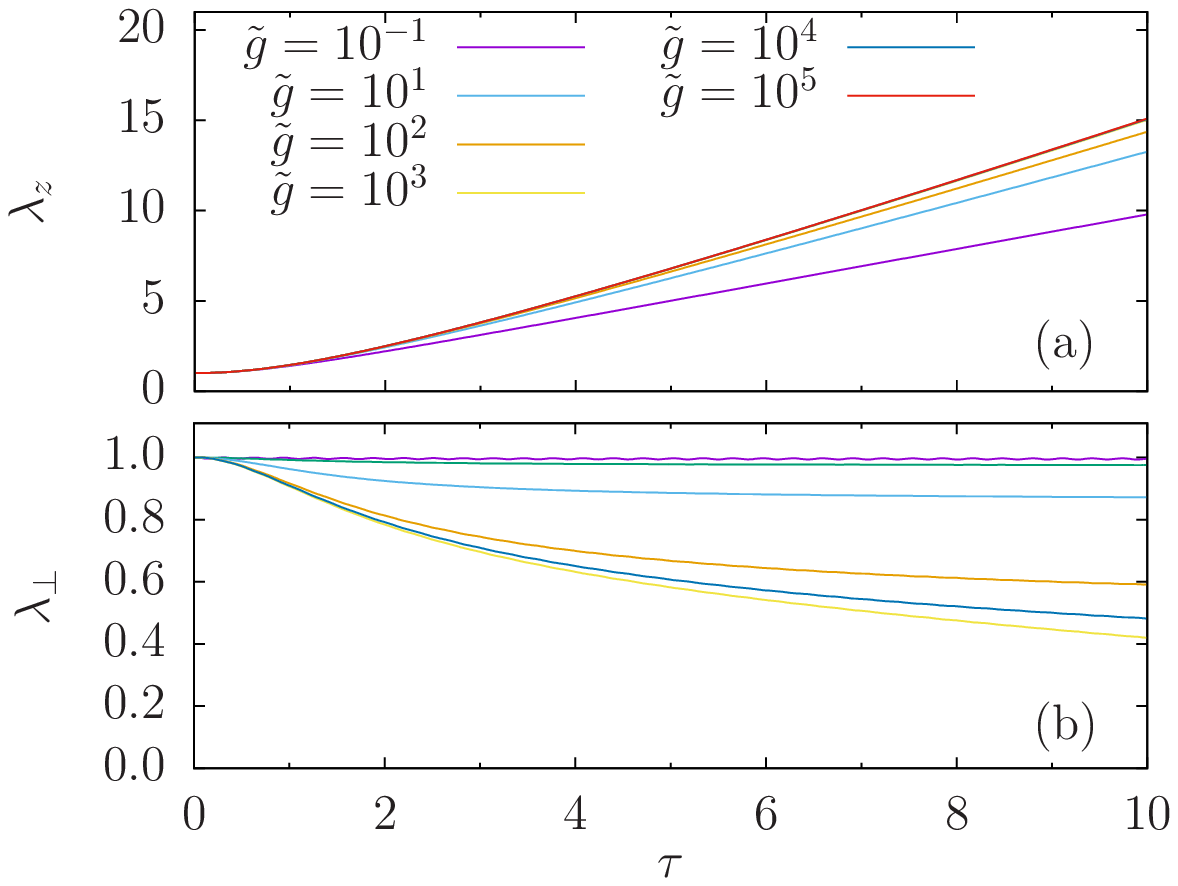}
 \hspace{1cm}
 \includegraphics[width=0.9\columnwidth]{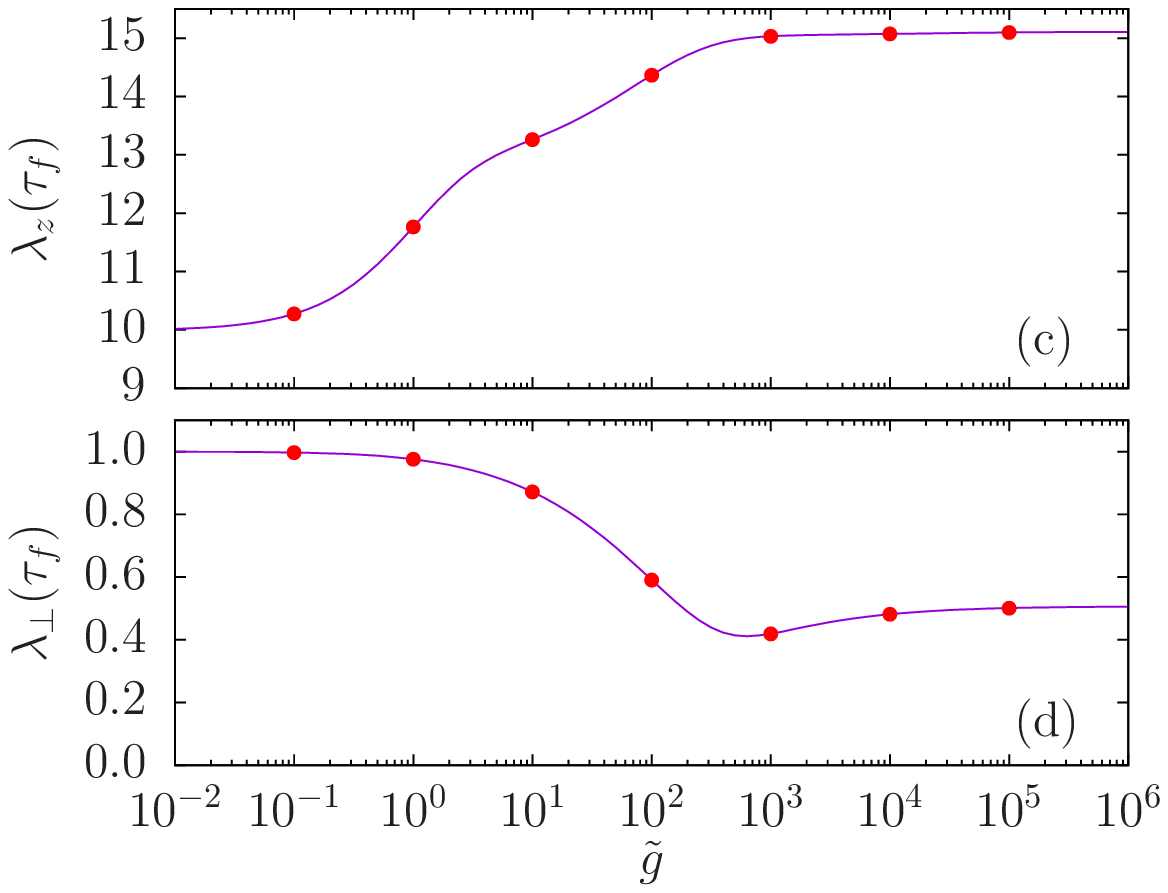}
\caption{Expansion in a waveguide: Behavior of [(a) and (b)] the axial and radial scaling parameters as a function of time for different interaction strengths, and [(c) and (d)] of the corresponding values at $\tau_f=10$ as a function of the interaction strength $\tilde{g}$. The red dots in (c) and (d) correspond to the values of $\tilde{g}$ considered in (a) and (b).}
\label{fig:lambda-cyl}
\end{figure*}

\begin{figure}
\centerline{
 \includegraphics[width=0.49\columnwidth]{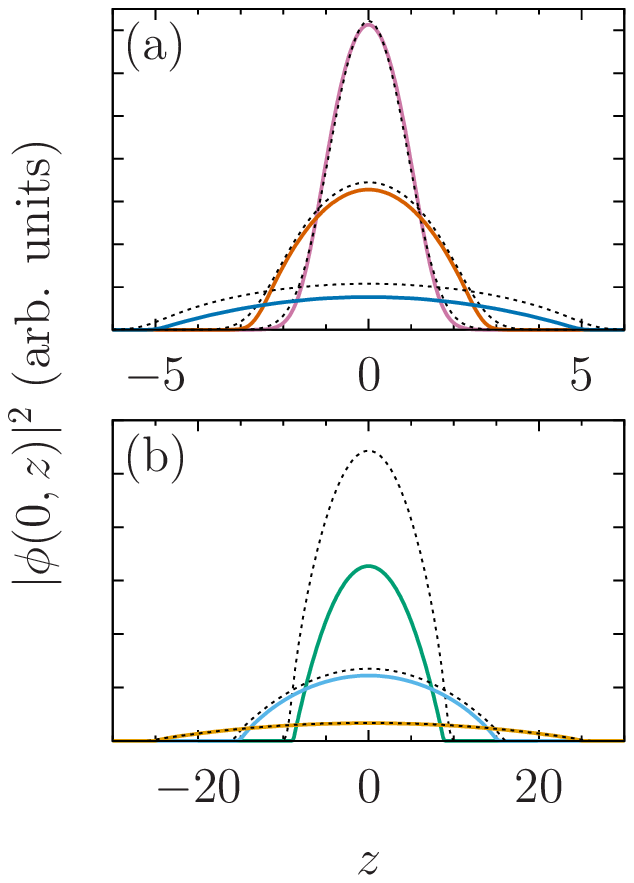}
 \includegraphics[width=0.49\columnwidth]{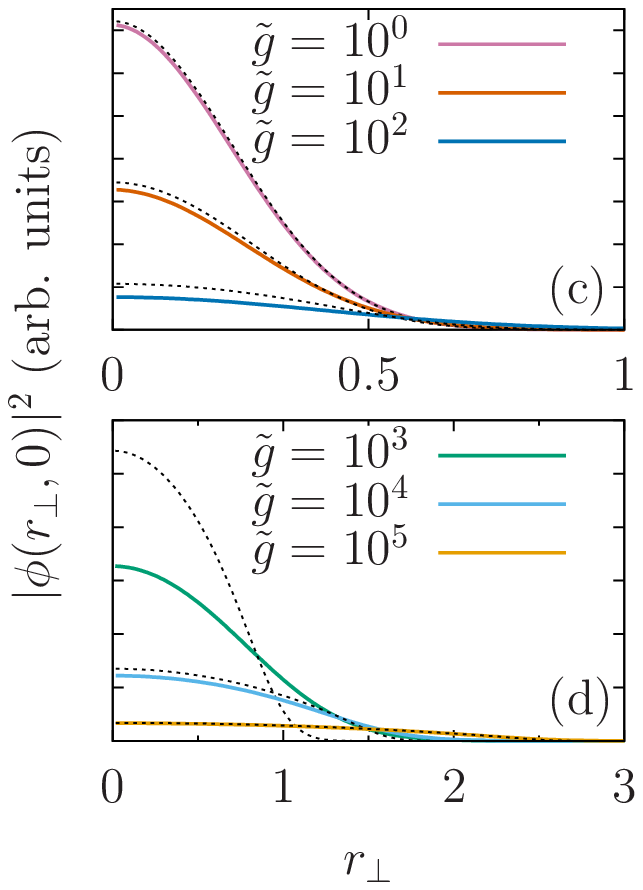}
 }
\caption{Expansion in a waveguide: Comparison of the density profiles as given by the numerical solution of the GP equation (solid) and by the scaling approach (dashed) for different interaction strengths and an evolution time of $\tau=10$. Density cuts along the $z$ axis at $r_\perp=0$ (left), and along the $r_\perp$ axis at $z=0$ (right). For intermediate values of $\tilde{g}$, the scaling solution largely overestimates  the central density and underestimates the radial width. The fidelity corresponding to each value of $\tilde{g}$ considered here is marked by the black circles in Fig.~\ref{fig:fidelity-cyl}.}
\label{fig:psi-cyl}
\end{figure}

Let us now turn to the case of a condensate initially confined by the same cylindrically symmetric potential in Eq.~(\ref{eq:cylpot}),
that is now allowed to expand in the presence of the radial confinement, following the switch-off of the axial potential [$\omega_z(\tau)=0$ for $\tau>0$]. 
This case represents a generalization of the quasi-1D case considered in Ref.~\cite{modugno2018}, the difference being that here the condensate profile along the radial direction is not restricted to that of the harmonic oscillator ground state -- several transverse excited states can be occupied, depending on the values of $\tilde{g}$ and $\alpha$. The 1D mean-field limit of Ref.~\cite{modugno2018} is recovered for $\tilde{g} \ll 4\pi\sqrt{\alpha}$, see Ref.~\cite{menotti2002b}. 

Considering that the expansion takes place along the axial direction, in this section we express frequencies in units of the axial frequency, namely 
$\omega_z(0)=1$, $\omega_{\perp}(0)\equiv\alpha$ \footnote{This choice corresponds to having dimensionless quantities in terms of the axial oscillator units. In this case $\tilde{g}\equiv 4\pi a/a_{z}$.}. In the following we shall use $\alpha=10$, as an example.
In the present case, the self-similar ansatz for the density takes the same form as in Eq.~(\ref{eq:scaling-cyl}), with Eq.~(\ref{eq:lrf}) being replaced by
\begin{equation}
\ddot{\lambda}_{\perp} = \frac{A_1(\tilde{g})}{\lambda_{\perp}^3} + \frac{A_2(\tilde{g})}{\lambda_{\perp}\lambda_z^2} + \frac{A_3(\tilde{g})}{\lambda_{\perp}^3\lambda_z} -\alpha^2 \lambda_{\perp}, 
\end{equation}
where the constants $A_i$ and $B_i$ are those previously defined (modulo a straightforward rescaling, owing to the different choice of $\omega_{0}$).

The behavior of the scaling parameters $\lambda_z$ and $\lambda_{\perp} $as a function of time (for different values of $\tilde{g}$) and as a function of $\tilde{g}$ (after an expansion time $\tau=10$) are shown in Fig.~\ref{fig:lambda-cyl}, in the left and right panels respectively. 
Their behavior corresponds to a contraction of the radial size, and to a dilatation in the axial direction, as one would expect for a condensate expanding along the waveguide. Similarly, the monotonic behavior of $\lambda_z(\tau_{f})$ as a function of $\tilde{g}$ -- which again interpolates between the the two limiting behaviors (noninteracting, up to $\tilde{g} \approx \num{e-2}$, and TF for $\tilde{g} \gtrsim \num{e3}$) --, agrees with the naive expectations. 
Instead, the local minimum displayed in the behavior of $\lambda_{\perp}(\tau_{f})$ as a function of $\tilde{g}$, see Fig.~\ref{fig:lambda-cyl}(d), signals a failure of the self-similarity assumption, since the radial size is expected to decrease monotonically by increasing $\tilde{g}$.

Indeed, by looking at the density cuts along the axial and radial directions in Fig.~\ref{fig:psi-cyl}, it is manifestly evident that for intermediate values of the interactions the effective scaling approach breaks down. 
In these figures, the dashed profile corresponds to the initial state -- which is equivalent to the self-similar solution in the rescaled coordinate system --, and the solid line is the solution of the scaled GP equation at $\tau_{f} = 10$. It is evident that for intermediate values of $\tilde{g}$, where the radial scaling parameter becomes too small (see Fig.~\ref{fig:lambda-cyl}), the actual density profile is characterized by a strong contraction along the axial direction (and by a corresponding radial broadening) with respect to what a perfect scaling would predict (dashed lines in the figure).
As anticipated in the introduction, this behavior is due to the fact that self-similarity is explicitly violated in the presence of a residual trapping. In fact, the density profile along the trapped directions is determined by the interplay between the confining potential and the nonlinear term, and it therefore gets modified as the condensate expands: It cannot be maintained during the expansion along the waveguide \footnote{We have also verified that in the case of a radial expansion in the presence of a fixed axial confinement (that is, a planar waveguide) the scaling approximation can become even worse, due to the fast decrease of the local density (the condensate is now allowed to expand along two spatial directions).}. In this respect, also in the case of $\tilde{g}>10^{4}$ -- for which the scaling approach appears reasonably accurate, see Figs.~\ref{fig:lambda-cyl} and  \ref{fig:psi-cyl} --, we expect this approach to eventually break down in the asymptotic regime of very low densities \footnote{Notice that if one were to trap the condensate again once it has reached the asymptotic, extremely diluted regime, the corresponding equilibrium profile would be a Gaussian.}.

\begin{figure}
\centering
 \includegraphics[width=0.9\columnwidth]{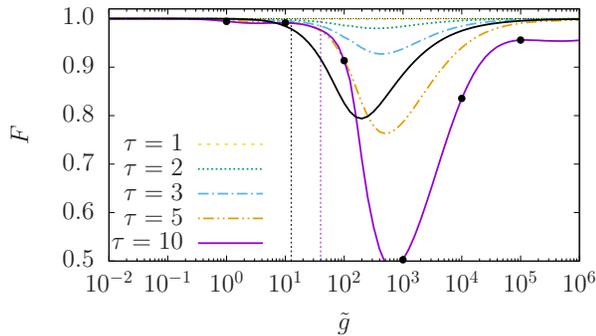}
\caption{Expansion in a waveguide: Plot of the fidelity as a function of the interaction strength, at different evolution times (for $\alpha=10$). The solid black line corresponds to the case with $\alpha=1$, at $\tau=\tau_f=10$. The vertical dashed lines represent the value $4\pi\sqrt{\alpha}$ for $\alpha=1,10$ (from left to right, respectively -- see text). The black circles correspond to the density profiles shown in Fig.~\ref{fig:psi-cyl}}
\label{fig:fidelity-cyl}
\end{figure}
Indeed, the above behavior can be inferred by looking at the fidelity, which is shown in Fig.~\ref{fig:fidelity-cyl} as a function of $\tilde{g}$, at different evolution times. 
As a reference, here we also show the value of the fidelity at $\tau=\tau_f$ for the case of a shallower radial confinement, namely $\alpha=1$. Then, by comparing the two cases at $\tau=\tau_f=10$, for $\alpha=1$ and $10$ (the black and yellow continuous lines, respectively), it is clear that in the former case the system is able to evolve self-similarly even for large values of $\tilde{g}$ (in the time range considered here), whereas this is not the case for $\alpha=10$. 
The reason behind this is that for a lower initial local density, the system takes longer to exit
the self-similar regime (for values of $\tilde{g}$ for which the scaling is exact in the quasi-1D limit). 
Consequently, we expect the self-similarity to also break down eventually for $\alpha=1$, though at larger expansion times than those in the case of $\alpha=10$.

Therefore, we can conclude that the effective scaling approach to the GP equation for the waveguide expansion works only if one of the following two conditions is satisfied: (i) short evolution times, when the radial profile has not yet deviated from the initial one (that is, the local density has not decreased too much) or (ii) the system is in the 1D mean-field limit \cite{menotti2002b,modugno2018}, $\tilde{g} \ll 4\pi\sqrt{\alpha}$, where the density profile is always characterized by a Gaussian transverse profile (for arbitrary times).

\section{Conclusions}

We have compared the exact evolution of an expanding 3D Bose-Einstein condensate with that obtained by means of an effective scaling approach based on the assumption of a self-similar dynamics. This approach consists in looking for self-similar solutions to be satisfied on average, by integrating the hydrodynamic equations over the spatial coordinates \cite{guery-odelin2002,menotti2002,hu2003,modugno2018}. Different geometries and configurations have been considered.

In case of isotropic traps, we have found that the hypothesis of self-similarity -- which is exact in the noninteracting and hydrodynamic limits \cite{dalfovo1999} -- is indeed rather accurate even for intermediate values of the interactions. This result provides an extension to three dimensions of the 1D proof-of-concept discussed in Ref.~\cite{modugno2018}. 
Conversely, we have found that significant deviations from the exact evolution characterize the expansion of prolate and oblate condensates. This behavior originates from the fact that when some of the directions expand much faster than the others, they produce local variations of the density that break the hypothesis of self-similarity.
Accordingly, we have found that self-similarity also breaks down for the expansion in a waveguide, due to the presence of the residual trapping. Indeed, the balance between the effect of the confining potential and that of the nonlinear term changes 
during the expansion owing to the local variations of the density. This in turn affects the density profile along the trapped directions, thus resulting in a behavior that is not self-similar. In the specific case of a cylindrical waveguide, we have shown that the effective scaling approach provides reliable results only for short evolution times, when the radial profile has not yet deviated from the initial profile (that is, when the local density has not decreased too much), or if the system is in the so-called 1D mean-field limit \cite{menotti2002b,modugno2018}, where the density profile is always characterized by a Gaussian transverse profile.

We also remark that although the hypothesis of self-similarity is justified only for certain geometries and configurations, the scaling approach can still be very useful for simulating a free expansion from the numerical point of view. Indeed, even if the scaling is only approximate, solving the dynamical equations in a rescaled coordinate system can greatly reduce the size of the computational grid. 
In this respect, the above analysis suggests that it could be useful to extend the present approach to the case of the time-of-flight expansion of binary mixtures as employed in current experiments, see, e.g., Refs. \cite{lee2018,burchianti2018}. Indeed, in some cases solving the equations for the expansion dynamics can be computationally prohibitive \cite{lee2018}, which is why scaling methods could be beneficial in this context. 
In addition, an effective self-similarity could be explored also for other expansion scenarios \cite{eckel2018}, and even for other equations.
One of the more relevant examples is in the context of quantum protocols, where scaling approaches can also be employed, see, e.g., Refs. \cite{schaff2011a,schaff2011b,torrontegui2013}. 

\acknowledgments
We acknowledge support from the Spanish Ministry of Science, Innovation and Universities and the European Regional Development Fund FEDER through Grant No. PGC2018-101355-B- I00 (MCIU/AEI/FEDER,UE), and the Basque Government through Grant No. IT986-16.

\appendix*
\section{Effective scaling for the GP equation}

\begin{figure}[b]
\centering
 \includegraphics[width=0.9\columnwidth]{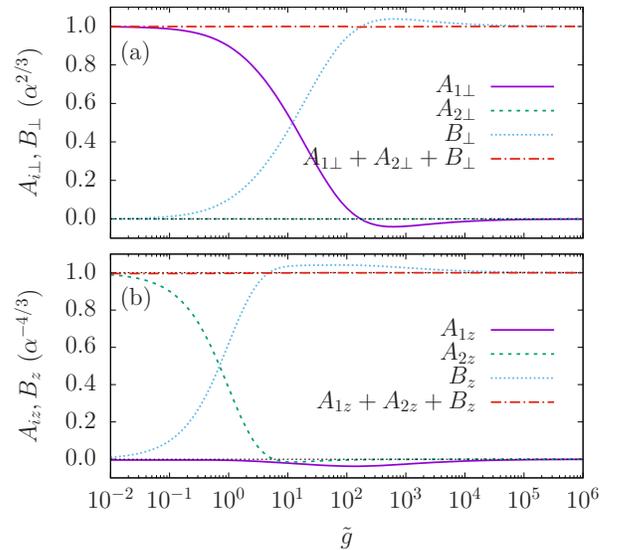}
 \caption{Behavior of the constants $A_{i\nu}$ and $B_{\nu}$ ($i=1,2$, $\nu=\perp,z$) as a function of $\tilde{g}$, for $\alpha=10$.}
\label{fig:AB-cyl}
\end{figure}

Here we derive the equations for the scaling parameters that characterize the approximate self-similar solutions of the Gross-Pitaevskii equation. For an axially symmetric system, Eq.~(\ref{eq:3-sc1}) consists of two equations, which take the following form ($\tilde{r}_{\perp} \equiv r_{\perp}/\lambda_{\perp}$, $\tilde{z} \equiv z/\lambda_z$):
\begin{widetext}
\begin{align}
\label{eq:app1}
\frac12\left(\ddot{\lambda}_{\perp} + \alpha^{2/3}\lambda_{\perp}\right)\lambda_{\perp}\tilde{r}^2_{\perp}+ \tilde{g}\frac{n_{0}(\tilde{r}_{\perp},\tilde{z})-n_{0}(0,z)}{\lambda_{\perp}^2\lambda_z} +
P(\tilde{r}_{\perp},\tilde{z};\tau) 
- \frac{K_{\perp}^{\perp}(\tilde{z})}{\lambda_{\perp}^2}- \frac{K_{z}^{\perp}(\tilde{z})}{\lambda_{z}^2} &= 0, 
\\
\frac12\left(\ddot{\lambda}_z + \alpha^{-4/3}\lambda_z\right)\lambda_z\tilde{z}^2 + \tilde{g}\frac{n_{0}(\tilde{r}_{\perp},z)-n_{0}(\tilde{r}_{\perp},0)}{\lambda_{\perp}^2\lambda_z} +
P(\tilde{r}_{\perp},\tilde{z};\tau) - \frac{K_{\perp}^{z}(\tilde{r}_{\perp})}{\lambda_{\perp}^2}- \frac{K_{z}^{z}(\tilde{r}_{\perp})}{\lambda_{z}^2}&= 0, 
\label{eq:app2}
\end{align}
with
\begin{equation}
P(\tilde{r}_{\perp},\tilde{z};\tau) = -\frac{1}{2\lambda_{\perp}^{2}(\tau)\sqrt{n_{0}}}\left(\frac{\partial^2}{\partial\tilde{r}_{\perp}^2} + \frac{2}{\tilde{r}_{\perp}}\frac{\partial}{\partial\tilde{r}_{\perp}}\right)\sqrt{n_{0}}
-\frac{1}{2\lambda_{z}^{2}(\tau)\sqrt{n_{0}}}
\frac{\partial^{2}}{\partial\tilde{z}^2}\sqrt{n_{0}}
\equiv P_{\perp}(\tilde{r}_{\perp},\tilde{z}) + P_z(\tilde{r}_{\perp},\tilde{z}),
\end{equation}
\end{widetext}
$K_{\nu}^{\perp}(\tilde{z}) \equiv \lambda_{\nu}^{2}P_{\nu}(0,\tilde{z})$, 
$K_{\nu}^{z}(\tilde{r}_{\perp}) \equiv \lambda_{\nu}^{2}P_{\nu}(\tilde{r}_{\perp},0)$, 
where $\nu=\perp,z$
(all the $K$'s being independent of time).
Then, we multiply Eqs. (\ref{eq:app1}) and (\ref{eq:app2}) by $n_{0}$ and we integrate over the volume to get rid of the spatial dependence. This yields Eqs. (\ref{eq:lrf}) and (\ref{eq:lzf}), where
\begin{align}
A_{1\nu}(\tilde{g}) &= {2(D_{\perp}^{\nu} - E_{k\perp})}/{\sigma_{\nu}^2}, 
\\
A_{2\perp}(\tilde{g}) &= {2(D_{z}^{\nu} - E_{kz})}/{\sigma_{\nu}^2}, 
\\
B_{\perp}(\tilde{g}) &= {2\tilde{g}(\bar{n}_{2}^{\nu} -\bar{n}_2)}/{\sigma_{\nu}^2}, 
\end{align}
with $\sigma_{\perp}^2=\langle r_{\perp}^2\rangle_0$, $\sigma_{z}^2=\langle z^2\rangle_0$,
$E_{k\nu}=-\frac12\langle \nabla^2_{\nu}\rangle_0$, 
\begin{align}
\bar{n}_{2} &= \int n_{0}^2(r_{\perp},z)dV, 
\\
\bar{n}_{2}^{\perp} &= \int n_{0}(0,z) n_{0}(r_{\perp},z) dV,
\\
\bar{n}_{2}^{z} &= \int n_{0}(r_{\perp},0) n_{0}(r_{\perp},z)dV,
\\
D_{\nu}^{\nu^{'}} &= \int n_{0}(r_{\perp},z) K_{\nu}^{\nu'}(z)dV, 
\end{align}
and $dV= 2\pi r_{\perp}dr_{\perp}dz$. An example of the behavior of the scaling parameters $A_{i\nu}$ and $B_{\nu}$ as a function of $\tilde{g}$, is shown in Fig. \ref{fig:AB-cyl}, for $\alpha=10$.

In case of spherical symmetry, the above expressions simplify to
\begin{align}
A(\tilde{g}) &= 2\left(D_{0} - E_k^0\right)/\sigma_r^2, 
\\
B(\tilde{g}) &= 2\left[\tilde{g}n_{0}(0) - 2E_{int}^0\right]/\sigma_r^2,
\end{align}
with (here $dV=4\pi r^2 dr$)
\begin{align}
\sigma_r^2 &= \int n_{0}(r){r}^2 dV \equiv\langle r^2\rangle_0,
\\
E_k^0 &= \frac12\int \left[\nabla_r \sqrt{n_{0}(r)}\right]^2 dV
\equiv-\frac12\langle \nabla^2_r\rangle_0,
\\
E_{int}^0 &= \frac{\tilde{g}}{2} \int n_{0}^2(r) dV.
\end{align}

Finally, we mention that the sum rules fulfilled by the constants $A$ and $B$ can be obtained as discussed in Ref. \cite{modugno2018}, via the stationary GP equation. For example, in the spherically symmetric case we have
$\left[-(1/2) \nabla^2 + (1/2)r^2 + \tilde{g}|\psi_0|^2\right] \psi_0 = \mu \psi_0$ that, by left-multiplying by $\psi_0=\sqrt{n_{0}}$, yields
$\mu= P(r,0) + (1/2)r^2 + \tilde{g}n_{0}$.
Then, by plugging the latter expression back into the GP equantio and integrating over the volume, one can easily obtain the sum rule $A+B=1$. Similarly, one can obtain the corresponding sum rules for the parameters $A_{i\nu}$ and $B_{\nu}$.

\bibliography{scaling}

\end{document}